# Antiferromagnetic phase of $MnTe_2$: spin-order-induced renormalization of phonon Raman spectrum

Taras Palasyuk[1*], Ashutosh S. Wadge[2], Eduard E. Zubov[3,4], Igor Antoniazzi[5], Gayatri[5], Andrzej Wiśniewski[2], Wojciech Szuszkiewicz[6], Daniel Jastrzebski[2], Ryszard Diduszko[6], Adam Babiński[5], Maciej R. Molas[5], Michał Wierzbicki[1] and Cezariusz Jastrzebski[1]

[1] *Faculty of Physics, Warsaw University of Technology, Koszykowa 75, Warsaw, PL-00662, Poland*
[2] *International Research Centre MagTop, Institute of Physics, Polish Academy of Sciences, Aleja Lotnikow 32/46, PL-02668 Warsaw, Poland*
[3] *Kyiv Academic University, 36 Acad. Vernadsky Blvd., UA-03142 Kyiv, Ukraine*
[4] *Donetsk Institute for Physics and Engineering named after O.O. Galkin, National Academy of Sciences, Kyiv, UA-03028, Ukraine*
[5] *Faculty of Physics, University of Warsaw, Pasteura 5, PL-02093 Warsaw, Poland*
[6] *Institute of Physics, Polish Academy of Sciences, Aleja Lotnikow 32/46, PL-02668 Warsaw, Poland*

Correspondence: Taras Palasyuk (taras.palasyuk@pw.edu.pl)

**ABSTRACT**

Recently, two modes of spin wave excitations (magnons) were detected in Raman scattering spectra of $MnTe_2$. It is an unconventional magnet that features a pyrite-type crystal structure hosting a non-collinear, non-coplanar antiferromagnetic (AFM) spin order below the Néel temperature. In this brief communication, the focus is primarily on the effect of spin ordering on lattice dynamics (phonons) that has been established through temperature (5 – 300 K), linear polarization angle ($0^o$ - $360^o$) and external magnetic field (up to 16 T) dependent measurements. It has been found that below the magnetic transition temperature, the frequencies of four out of five Raman-active lattice vibrations (i.e., $E_g$ and three $T_g$ phonon modes) revealed significant departure from the mechanism of the three-phonon anharmonic decay described by Klemens model. Here, we propose a new model considering spin-phonon coupling to explain frequency renormalization of the phonon part of the Raman spectrum upon the para-/antiferro- (PM/AFM) magnetic phase transition change in $MnTe_2$. Through employing Green's function formalism, we show that a simple two-sublattice collinear antiferromagnet approach follows well both spin ordering induced hardening and softening of phonon modes observed in AFM phase of $MnTe_2$ upon cooling. Additionally, the detailed temperature and field dependences of the energy for both magnon modes are presented. In addition to unexpected twofold anisotropy of the higher energy magnon mode (approx. 3.5 meV) unveiled in crossed configuration, our magneto-optical measurements revealed a subtle, yet definite softening trend for the lower energy mode (approx. 2.6 meV). Although the precise reasons for these effects remain unclear, we put forward a symmetry reduction in the local atomic surroundings as a possible mechanism for discussion.

## INTRODUCTION

The use of spin-group symmetry to classify magnetic materials has recently broadened the notion of what a magnet can be [1]. Beyond these conventional magnets, a growing family of unconventional magnets has emerged, including altermagnets [2],non-collinear spin-split antiferromagnets [3], and anomalous Hall antiferromagnets [4]. Although these systems carry no net magnetic moment, they exhibit phenomena once considered exclusive to ferromagnets, such as spin-split electronic bands and anomalous Hall responses. Among them, non-collinear, non-coplanar antiferromagnets are particularly intriguing, as their complex spin textures shape both the magnetic excitation spectrum and the way the lattice responds to magnetic ordering [5]. Magnetic materials with non-collinear spin arrangements thus provide a natural setting for investigating spin-phonon coupling (SPC), the mutual influence of magnetic order and lattice vibrations [6]. Alongside their magnon excitations, changes in the phonon spectrum offer a complementary view of how the lattice responds to magnetic ordering. A microscopic understanding of SPC in such complex magnets is therefore of both fundamental and practical interest, underpinning efforts in magnonic engineering and spintronics.

Until recently systematic investigations using Raman spectroscopy have been performed on two groups of compounds with collinear antiferromagnetic structures. The first group comes under the category of transition metal difluorides ($TMF_2$) such as $MnF_2$ [7], $FeF_2$ [7], $CoF_2$ [8] and $NiF_2$ ( [8], [9]), which were mostly investigated in 1980s. These materials share a 3D tetragonal rutile-type structure ($TiO_2$-type). Researchers used them to establish early models of how magnetic order affects lattice vibrations (phonons) in a rigid 3D framework [8], [10]. Whereas the second category known as transition metal phosphorus trichalcogenides ($TMPX_3$) started gaining wide interest since 2015. These are layered van der Waals materials where magnetic ordering can persist down to the monolayer limit, offering a completely different environment for spin-phonon coupling compared to 3D crystals [11]. Because $TMF_2$ is a tightly bound 3D ionic lattice and $TMPX_3$ is a loosely bound 2D layered structure with covalent p-p (phosphorous) bonds, the physics governing their spin-phonon interactions is distinct. This lack of a "middle ground" makes it difficult to create a unified theory that applies to all magnetic materials.

A natural test case for extending such studies beyond collinear magnets is $MnTe_2$ [12], [13] a three-dimensional cubic pyrite-type antiferromagnet whose magnetic ground state features a non-collinear, non-coplanar spin texture [14]. The compound is regaining considerable research attention as an emergent unconventional magnet, following the observation of momentum-dependent spin splitting in its electronic structure [14].

Here we investigate Raman scattering in the antiferromagnetic phase of $MnTe_2$. Recently, we unambiguously established the complete first-order phonon Raman spectrum of $MnTe_2$ in its paramagnetic (PM) phase [15] revealing the nearly degenerated $E_g$ and $T_g$ phonons modes. Below the Néel temperature, up to date, Raman scattering study of $MnTe_2$ are represented by the papers of Wulferding et al. [16], Uykur et al. [17] and a few non-peer-reviewed contributions (PhD thesis [18], two preprint reports [19], [20]). In addition to phonon-related signals above 90 $cm^{-1}$, these studies reveal distinct spectral features in the measured Raman spectrum that are identified as spin-wave excitations at 20 $cm^{-1}$ and 30 $cm^{-1}$. It is worth noting that in the report of Wulferding et al. [16] a pronounced circular Raman dichroism and anomalous rotational symmetry in the magnetic excitation spectrum, namely, for a spin-wave mode at 20 $cm^{-1}$ were revealed. None of the cited reports show the phonon spectrum variations across the PM–AFM transition that would indicate potential SPC in $MnTe_2$.

Our present report details major changes in the Raman-active phonon modes involving predominantly non-magnetic (tellurium) atoms induced by progressive ordering of spins on magnetic (manganese) atoms. We show that basic phonon properties experience considerable renormalization upon paramagnetic-antiferromagnetic phase transition. Here, we develop a new model similar to the one used by Cottam and Lockwood [8] for simple collinear rutile-type AFM $TMF_2$ that explains the phonon frequency renormalization as a result of spin-spin correlations. In this model spin and phonon Green's functions are studied within the zeroth approximation of the effective self-consistent field for a two-sublattice antiferromagnet. Spin-phonon coupling is realized through the action of the vibrational potential, which modulates the exchange interactions between magnetic ions. Hence, the temperature dependence of the phonon modes is attributed to both magnetic ordering and anharmonic contributions.

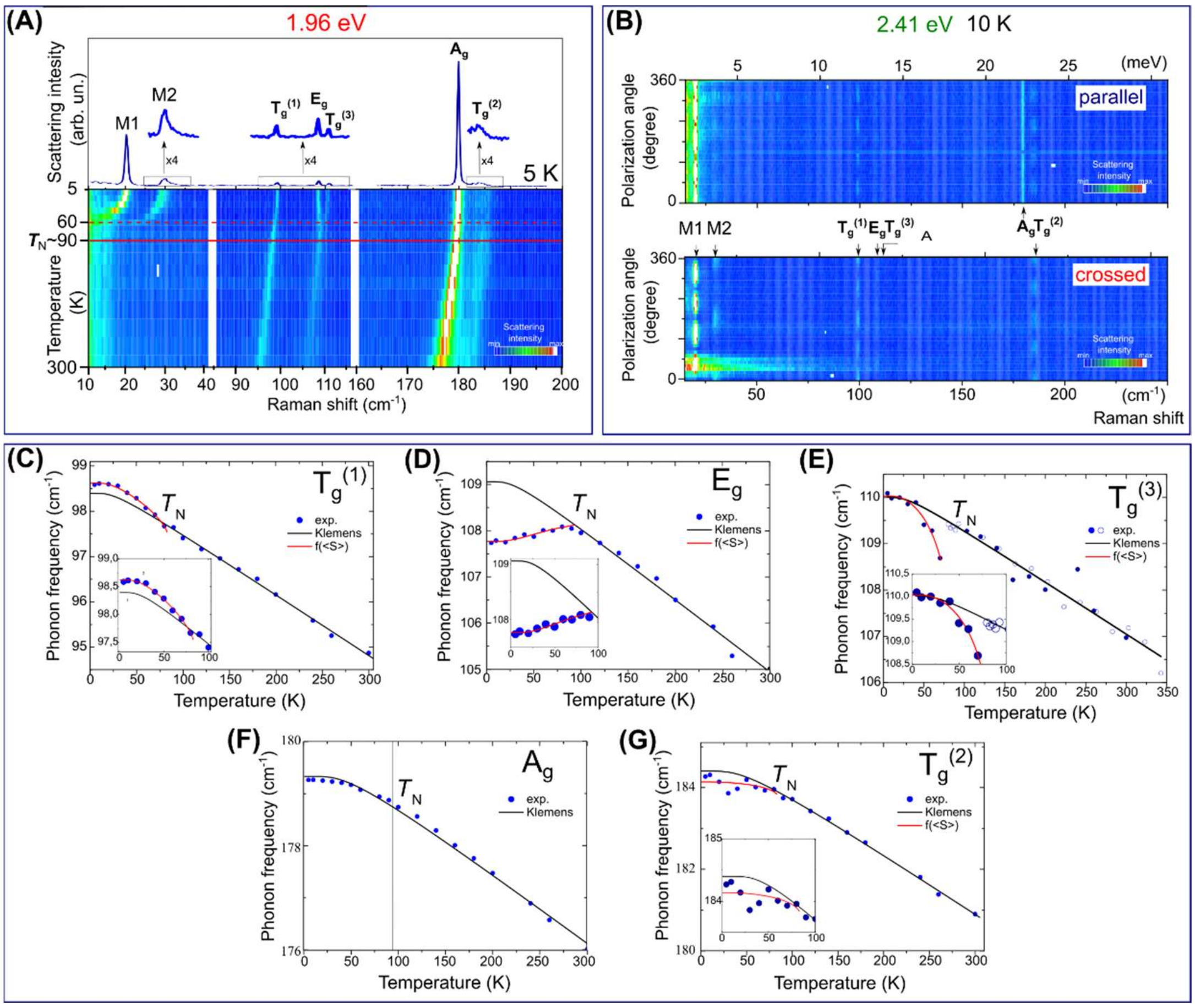

**Figure 1. Investigation of the effect of temperature on Raman scattering in $MnTe_2$. (A)** Evolution of nonpolarized Raman scattering spectra collected in the range 5 – 300 K using excitation 1.96 eV. "$T_N$" – the Néel temperature. Lattice modes (phonons) are denoted as $T_g^{(1)}$, $E_g$, $T_g^{(3)}$, $A_g$ and $T_g^{(2)}$ spectral bands. Below 60 K the progressive appearance of new signals (M1, M2) in the ultra-low frequency range was observed. Detailed view on the spectrum acquired at 5 K **(B)** Intensity of Raman scattering as a function of polarization angle measured with parallel (VV) and crossed (VH) configurations at 10 K using 2.41 eV excitation. Obtained results confirmed symmetries for all five Raman-active phonon modes. **(C) – (G)** Detailed analysis of temperature dependences of the $T_g^{(1)}$, $E_g$, $T_g^{(3)}$, $A_g$, and $T_g^{(2)}$ phonon frequencies taken from nonpolarized Raman spectra. Experimental data from this study are shown as blue filled circle symbols whereas open circles denote data from Ref. [15]. Klemens model fit (solid black lines) to the experimental data of the paramagnetic (PM) phase ($< S^z >_0 = 0$). Solid red lines represent the fits using the Equation (4) (see in the text) for the antiferromagnetic (AFM) phase data. Insets show an enlarged view of the temperature range below the Néel temperature.

## EXPERIMENTAL RESULTS.

Overall, two series of temperature-dependent measurements were performed on single-crystal samples from different synthesis batches yielding highly consistent results.

**Figure 1 (A)** presents Raman scattering spectra from $MnTe_2$ acquired in the temperature range 5 – 300 K, which clearly reveal all five Raman-active lattice vibration (phonon) modes. Polarized Raman scattering measurements (as shown in **Figure 1(B)** and **Table 1**) performed at 10 K confirmed that the symmetry of phonon modes, i.e. $T_g^{(1)}$, $E_g$, $T_g^{(3)}$, $A_g$ and $T_g^{(2)}$ (in order of increasing frequency) remain unaltered across the paramagnetic-to-antiferromagnetic phase transition (occurring at ca. 90 K). These results are consistent with those recently reported by Wulferding et al. [16]. As regards temperature dependence of phonon frequency, in the range of stability of PM phase, i.e. approx. 90 – 300 K, all phonon modes closely followed the temperature dependence predicted by the Klemens model (three-phonon anharmonic decay mechanism), in agreement with our recent study [15] (**Figure 1 (C) – (G)**). In contrast, upon further cooling below the Néel temperature ($T_N$), noticeable departures of the frequencies from their temperature dependencies expected for solely phonon-phonon relaxation, were observable for three $T_g$ and $E_g$ modes. Specifically, an abrupt blueshift (frequency increase) and distinct redshift (frequency decrease) were noted for the lowest $T_g^{(1)}$ and the highest $T_g^{(2)}$ lying modes, respectively (**Figure 1 (C) – (G)**).

**Table 1.** Raman intensity at 10 K as a function of the angle between incident and scattered light polarization. '*a*', '*b*' and '*d*' – are tensor elements; φ - offset angle.

| Mode frequency ($cm^{-1}$) magnon | Mode frequency ($cm^{-1}$) phonon | Symmetry | Intensity relation: Parallel (VV) | Intensity relation: Crossed (VH) |
|---|---|---|---|---|
| | 180 | $A_g$ | $I \propto a^2$ | 0 |
| 20.6 | 108 | $E_g$ | $I \propto b^2(1 + 3\cos^2 2(\theta + \varphi))$ | $b^2 \sin^2 2(\theta + \varphi)$ |
| 28* | 99<br>108<br>183 | $T_g$ | $I \propto d^2 \sin^2 2(\vartheta + \varphi)$ | $d^2 \cos^2 2(\theta + \varphi)$ |

* shows twofold anisotropy

The most pronounced deviations from the Klemens model were revealed for neighboring $E_g$ and $T_g^{(3)}$ modes. Due to proximity of their energies, they remained poorly resolved in the spectra collected in the temperature range approx. 90 – 300 K and were not resolved in the range 60 – 90 K. Instead, below 60 K the phonon mode $T_g^{(3)}$ started to emerge as a shoulder of $E_g$ at its high frequency side, further evolving in a clearly separated Lorentzian peak. Nonetheless, having mode symmetries well established in both phases and having combined data from this study with the data for the PM phase from our recent study [15], we could reliably track temperature dependencies of $E_g$ and $T_g^{(3)}$ phonons in the AFM phase as presented in the corresponding plot of **Figure 1 (E)**.

With decreasing temperature below $T_N$ to 5 K, it was found that the frequency of the $E_g$ mode decreased (softened) in a linear manner yielding the overall drop of approx. 1.3 $cm^{-1}$, if compared to expected value of the Klemens model (**Figure 1 (D)**). The $T_g^{(3)}$ mode exhibits a more complex temperature dependence, showing an abrupt drop of approx. 1 $cm^{-1}$, when cooled from 80 K to 70 K, and, with subsequent lowering temperature below approx. 70 K, gradual increase resembling that of order parameter-like behavior. Similar observations regarding the frequency change of the two above-mentioned neighboring signals were reported in the Ref. [18]. However, several experimental factors—including a limited temperature range (< 120 K), sparse probing near the Néel temperature (every 20 K), and weak overall spectral intensity—prevented obtaining complete symmetry information for all measured signals. Consequently, the progressive separation of two signals with different symmetries was misinterpreted solely as a splitting of the $E_g$ mode.

At the same time the totally symmetric $A_g$ mode showed gradual increase (blueshift) of its frequency consistent with the Klemens model (**Figure 1 (F)**).

More detailed analysis on temperature dependencies of all Raman-active phonon modes will be given in the following Section "**RESULTS OF THEORETICAL MODELLING**".

## RESULTS OF THEORETICAL MODELLING.

The experimental results suggest that the modifications of decay mechanism (departure from Klemens model) found for phonon modes in the AFM phase may be coupled to the ordering of magnetic spins. Temperature-dependent Raman studies of spin–phonon coupling have a long history [7], [21], [22], [23], [24]. Specifically, a spin-phonon coupling was considered as a result of modulation of the superexchange $J_{ij}$ between the nearest *i* and *j* magnetic ions by the phonon excitations from non-magnetic sublattice [7], [22]. The series expansion for $J_{ij}$ over displacement $\Delta r_{ij}=r_i - r_j$ was made up to second order. In the first order of pointed expansion the detailed picture of spin-wave spectrum in a Heisenberg antiferromagnetic with spin-phonon coupling is studied in work [23]. A similar model of the influence of antiferromagnetic spin ordering on the far-infrared active optical phonon modes was presented in work [24] in which the low-temperature representations for spin operators are used to calculate the frequency renormalization. Further, it is supposed that spin correlations are independent of $\Delta r_{ij}$ because the spin relaxation to equilibrium is not achieved during the time of a lattice vibration [22].

The experimental temperature dependences of the phonon modes $E_g$ and $T_g$ below V temperature show an appearance of the ordered state. It is supposed that this feature is caused by anharmonicity of second order in spin-phonon interaction, which for the general case has a form [21]:

$$\hat{H}_{SP2} = -\sum_{i,\alpha,\boldsymbol{q}_1,\lambda_1,\boldsymbol{q}_2,\lambda_2} B(\boldsymbol{r}_{i\alpha},\boldsymbol{q}_1\lambda_1,\boldsymbol{q}_2\lambda_2) K_{12}(\boldsymbol{r}_i-\boldsymbol{r}_\alpha)\Phi_{\boldsymbol{q}_1\lambda_1}\Phi^+_{\boldsymbol{q}_2\lambda_2} \quad . \tag{1}$$

Here, $B(\boldsymbol{r}_{i\alpha},\boldsymbol{q}_1\lambda_1,\boldsymbol{q}_2\lambda_2)$ is the 2nd order constant of spin-phonon binding, spin correlator $K_{12}(\boldsymbol{r}_i-\boldsymbol{r}_\alpha) = < S_i^+S_\alpha^- + S_i^-S_\alpha^+ >$ and $\Phi_{\boldsymbol{q}\lambda} = b_{\boldsymbol{q}\lambda} + b^+_{-\boldsymbol{q}\lambda}$, where $b^+_{\boldsymbol{q}\lambda}$ ($b_{\boldsymbol{q}\lambda}$) are corresponding creation (annihilation) phonon operators with wave vector $\boldsymbol{q}$ and $\lambda$ polarization.

To find the phonon Green's function $U_\lambda(\tau,\boldsymbol{q}) = -< T_\tau \Phi_{\boldsymbol{q}\lambda}(\tau)\Phi^+_{\boldsymbol{q}\lambda}(0)>$ one can use the Dyson equation with mass-operator $\frac{1}{N}\sum_{\boldsymbol{k}} B(\boldsymbol{k},\boldsymbol{q}\lambda,\boldsymbol{q}\lambda)K(-\boldsymbol{k})$, where $B(\boldsymbol{k},\boldsymbol{q}\lambda,\boldsymbol{q}\lambda)$ and $K(\boldsymbol{k})$ are Fourier transforms of $B(\boldsymbol{r}_{i\alpha},\boldsymbol{q}_1\lambda_1,\boldsymbol{q}_2\lambda_2)$ at $\boldsymbol{q}_1=\boldsymbol{q}_2=\boldsymbol{q}$ and $K_{12}(\boldsymbol{r}_i-\boldsymbol{r}_\alpha)$, respectively. Then it follows the next expression for Fourier transform of $U_\lambda(\tau,\boldsymbol{q})$:

$$U_\lambda(i\omega_n,\boldsymbol{q}) = \frac{U_{0\lambda}(i\omega_n,\boldsymbol{q})}{1-\beta U_{0\lambda}(i\omega_n,\boldsymbol{q})\frac{1}{N}\sum_{\boldsymbol{k}} B(\boldsymbol{k},\boldsymbol{q}\lambda,\boldsymbol{q}\lambda)K(-\boldsymbol{k})} \quad , \tag{2}$$

where $U_{0\lambda}(i\omega_n,\boldsymbol{q}) = \frac{2\Omega_{q\lambda}}{\beta(\Omega^2_{q\lambda}-(i\omega_n)^2)}$ is unperturbed phonon Green's function, the discrete complex frequency parameter $i\omega_n = 2i\pi n/\beta$ *for* Bose subsystem, $1/\beta = T$ is temperature and $\Omega_{q\lambda}$ is phonon frequency. The pole of $U_\lambda(i\omega_n,\boldsymbol{q})$ after analytical continuation $i\omega_n \to \omega + i\delta$ gives the spectrum of phonon excitations $\omega_{q\lambda}$ with spin-phonon interactions, which is determined from equation

$$\tilde{\omega}_{q\lambda} = \sqrt{\Omega^2_{q\lambda} - 2\Omega_{q\lambda}\frac{1}{N}\sum_{\boldsymbol{k}} B(\boldsymbol{k},\boldsymbol{q}\lambda,\boldsymbol{q}\lambda)K(-\boldsymbol{k})} \approx \Omega_{q\lambda} - \frac{1}{N}\sum_{\boldsymbol{k}} B(\boldsymbol{k},\boldsymbol{q}\lambda,\boldsymbol{q}\lambda)K(-\boldsymbol{k}) \quad , \tag{3}$$

Here, it is supposed that $B(\boldsymbol{k},\boldsymbol{q}\lambda,\boldsymbol{q}\lambda) << \Omega_{q\lambda}$.

Calculating the spin correlator $K_{12}(\boldsymbol{r}_i-\boldsymbol{r}_\alpha)$ is rather a complex task. Within the framework of a two-spin cluster approximation, $K_{12}(\boldsymbol{r}_i-\boldsymbol{r}_\alpha)$ has been calculated in Ref. [8] and for $T \to T_N$ it is equal to $-S(S+1)/(Z_0-1)$, where $Z_0$ denotes the number of nearest neighbors on the opposite sublattice. In the Bogolyubov-Tyablikov approximation, this correlator is proportional to the average spin $< S^z >_0$ [25]. The present theory based on zeroth approximation of effective field also gives $K_{12}(\boldsymbol{r}_i-\boldsymbol{r}_\alpha) \sim < S^z >_0$. Thus, considering the phonon-phonon anharmonicity [22] and Eq. (3) finally one can write the Raman spectrum $\omega_{q\lambda}$ in general case as

$$\omega_{q\lambda} \approx \Omega_{q\lambda} + C\frac{< S^z >_0}{\Omega_{q\lambda}} - G\left(1+\frac{2}{e^{\left(\frac{\Omega_{q\lambda}}{2T}\right)}-1}\right) \tag{4}$$

with fitting parameters $\Omega_{q\lambda}$, *C* and *G*. The second term in Eq. (4) is inversely proportional to $\Omega_{q\lambda}$ since it follows from definition of Fourier transform of constant of spin-phonon binding $B(\boldsymbol{k},\boldsymbol{q}\lambda,\boldsymbol{q}\lambda)$ [21].

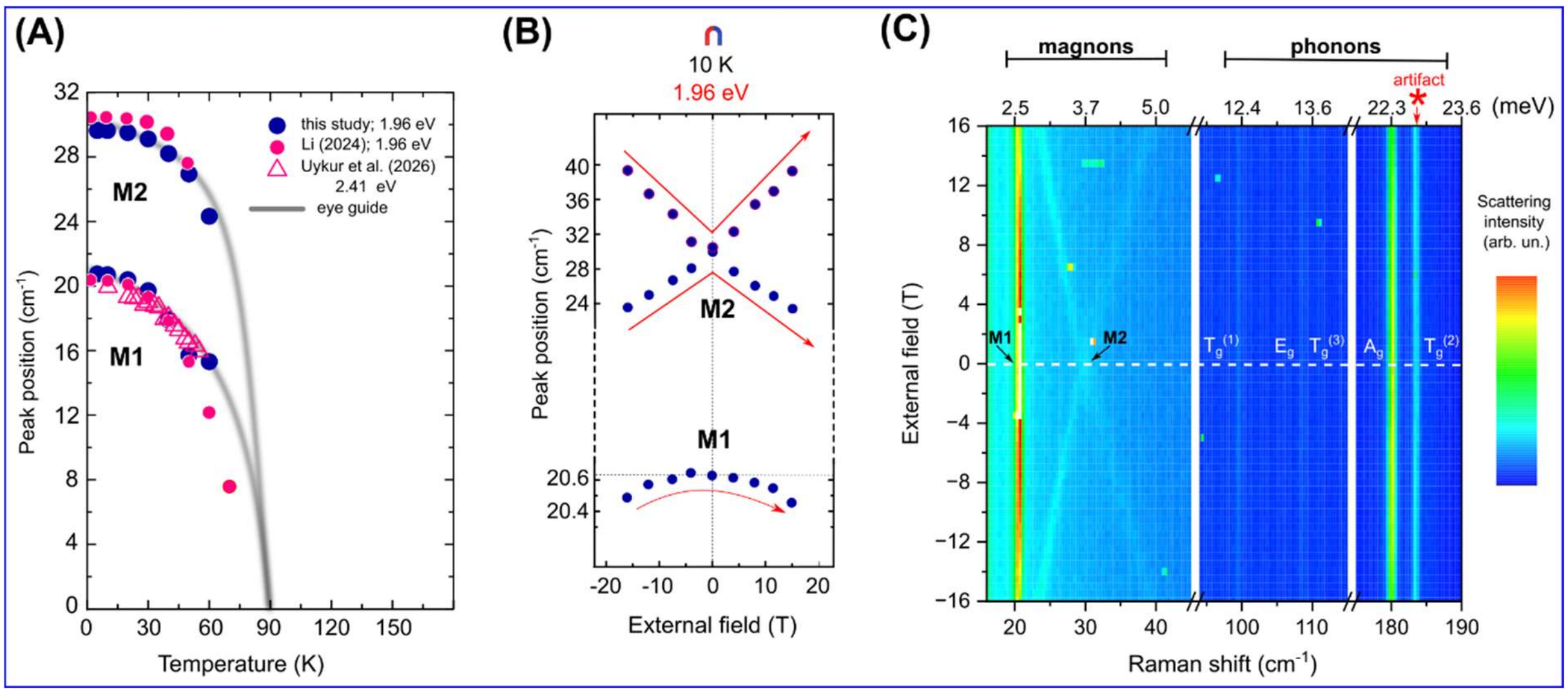

**Figure 2. Temperature and magnetic field dependence of the ultra-low frequency signals (M1, M2) studied by nonpolarized Raman scattering measurements in $MnTe_2$. (A)** Temperature dependences of M1 and M2 peak positions measured in this study (blue filled circles) and compared with available literature data. **(B)** Field dependence of M1 and M2 peak positions. **(C)** 2D contour plot showing the evolution of the Raman scattering spectra in applied magnetic field -16 to 16 T. Asterisks denote artifacts originating from the measurement setup.

The values of parameters extracted from fitting Equation (4) to the experimental data are presented in **Table 2**.

**Table 2.** Values of parameters $\Omega_{q\lambda}$, *C* and *G* derived from fitting Equation (4) to experimental data.

| | $T_g^1$ | | $E_g$ | | $T_g^3$ | | $A_g$ | $T_g^2$ | |
|---|---|---|---|---|---|---|---|---|---|
| | PM | AFM | PM | AFM | PM | AFM | PM, AFM | PM | AFM |
| $\Omega_{0\lambda}$, *cm⁻¹* | 98.9 | 98.3 | 109.7 | 107.3 | 109.9 | 105.4 | 180.2 | 185.4 | 183.6 |
| *G, cm⁻¹* | 0.48 | 0.37 | 0.62 | -0.33 | 0.46 | -0.38 | 0.87 | 0.98 | -0.10 |
| *C, (cm⁻¹)²* | 0 | 37.9 | 0 | 9.7 | 0 | 255 | 0 | 0 | 47.7 |

In parallel to the changes detected in the phonon part of the Raman spectrum, i.e. > 90 $cm^{-1}$, significant changes were also revealed in the ultra-low frequency range close to the laser line (**Figure 1 (A)**). Starting at 60 K, two broad features emerged at ca. 14 $cm^{-1}$ and 23 $cm^{-1}$. With further cooling, they gained considerably in intensity evolving in well-defined strong and medium signals denoted as M1 and M2, respectively. Intensity profiles measured in parallel configuration of angle resolved polarized Raman measurements confirmed fourfold anisotropy with $E_g$ and $T_g$ symmetry for M1 and M2, respectively. In contrast, for crossed configuration we noted twofold anisotropy of M2 whereas the anisotropy of M1 remained unchanged (**Figure 1 (B)** and **Table 1**). Similar observations were reported in Ref. [16].

In **Figure 2** we analyze ultra-low frequency spectral range, where scattering signals from spin wave excitations appear. We first compare temperature dependence measured in this study with experimental data from Refs. [18], [19] (**Figure 2 (A)**). For both modes, the plotted data demonstrate a consistent evolution characteristic of an order parameter.
Results of Raman scattering measurements under external magnetic field performed at 10 K revealed Zeeman splitting of the M2 signal confirming its spin-wave origin **Figure 2 (B), (C)**. Similar observations were reported in Refs [18], [19]. At the same time a small but systematic decrease (softening) of the M1 frequency (approx. 0.2 $cm^{-1}$) was detected in applied magnetic field up to 16 T. According to theoretical analyses presented in Refs. [16], [18], [19] this signal might originate from the singular spin wave mode.

## DISCUSSION

The standard Klemens model accounts for an anharmonic phonon-phonon decay background, where cooling-induced lattice contraction causes a monotonic hardening (blueshift) of phonon energies. Here, we compare the temperature dependence of phonon frequencies observed in $MnTe_2$ with those of $MnF_2$ [7] and related transition metal difluorides [8], for which systematic temperature-dependent Raman scattering investigations have been reported. The latter compounds crystallize in the rutile structure, which hosts four Raman-allowed lattice vibrations. Within their antiferromagnetic phases, all the Raman-active phonon modes exhibit frequency deviations from Klemens-model predictions, manifesting as either mode softening or hardening ranging from 0.13% to 2.8%. Comparing the absolute frequencies at the base temperature (ca. 10 K), the specific deviations for the $MnF_2$ modes are 0.6% ($A_{1g}$), 2.8% ($B_{1g}$), 0.3% ($B_{2g}$), and 0.5% ($E_g$). This spin-phonon coupling effect is of a comparable magnitude in $MnTe_2$, accounting for approximately 1% of the phonon energy for the $E_g$ mode, whereas it remains slightly weaker – on the order of 0.1% – for both the $T_g^{(1)}$ and $T_g^{(2)}$ modes. One can say that the contributions of second order anharmonicities in spin-phonon and phonon-phonon interactions depend on symmetry of phonon mode. In particular, for $A_g$ mode the spin-phonon binding is absent. Evidently, it relates to magnetic order, which has the same $A_g$ symmetry.

The temperature behavior of $T_g^{(3)}$ and $E_g$ modes in $MnTe_2$ requires a further comment. While a comprehensive explanation for these observations remains unclear, we propose the following tentative frameworks for consideration. The first observation is related to a sudden change in the energy of $T_g^{(3)}$ noting an abrupt drop by ca. 1 $cm^{-1}$ in a narrow temperature range on cooling from, approx. 80 K to 70 K, followed by a gradual blueshift (energy increase by ca. 1 $cm^{-1}$) with further cooling to 5 K. The second observation is that unlike other modes, the $E_g$ mode was noted to decrease in a monotonic and linear manner, by ca. 0.25 $cm^{-1}$ of its energy. In fact, these two modes exhibited the highest sensitivity to internal AFM order. Together, the effect of temperature on $T_g^{(3)}$ and $E_g$ resembles the one observed in the antiferromagnetic phases of $MnPS_3$, [26] $MnPSe_3$ [27], $FePS_3$ [28], ascribed to magnon-phonon hybridization driven anticrossing behavior of specific phonon modes and two magnons excitations. Indeed, the signal at higher energy of $T_g$ symmetry (ascribed in this study to $T_g^{(3)}$ phonon mode) does not split under applied magnetic field (**Figure 2 (C)**), which would also be valid for a two magnon excitation. This resemblance, however, seems to be coincidental since in pristine, undistorted cubic pyrite structure (space group Pa-3), $E_g$ and $T_g$ represent entirely distinct, orthogonally protected irreducible representations dictated by global symmetry. That would exclude the possibility of their mutual coupling. Nevertheless, such a scenario cannot be entirely ruled out, particularly if the antiferromagnetic ordering induces a reduction in crystal symmetry at the local level, even if the macroscopic global symmetry remains undistorted.

According to results of Mössbauer spectroscopy, presented in the earlier study of Kasai et al. [12], the angle between the Mn spin axis and the principal axis of the electric field gradient (EFG) changes rapidly between 4.2 K and 70 K, crossing an anomaly around 60 K, which is close to the turning point in the temperature dependence of the $T_g^{(3)}$ phonon mode found in this study. The EFG is highly sensitive to the local symmetry of the lattice surrounding. It was -proposed that variations in the internal positional parameter of the Te atoms couple to spin rotation and local coordination changes – specifically within the Mn polyhedra (Mn–Te bond lengths) and Te–Te dimers – without inducing a macroscopic structural phase transition.

This local structural adjustment may have implication to Raman selection rules favoring an increased probability of coupling between the $E_g$ and $T_g$ modes – both for phonons and magnons. At a localized level, this acts as a perturbation that removes the strict orthogonal protection between $E_g$ and $T_g$ representations. In view of the nearly degenerate energies of $E_g$ and $T_g$ phonon modes in $MnTe_2$ (ca. 1.5 $cm^{-1}$), even a tiny local symmetry-breaking perturbation may induce their hybridization.

The fact that the atomic parameter varies precisely alongside the rotation of the magnetic spin axis proves that the lattice and the magnetic subsystems are intricately locked. As the non-collinear spin structure continuously alters its local orientation below 70 K [12], the changing magnetic environment breaks time-reversal symmetry on the local scale as well. This may further enable coupling between magnon and phonon modes, including those at both the zone center and edge (e.g. from M – R points of the BZ). In this regard, the signal M2 showing clear asymmetry of its peak shape (5 K in **Figure 1 (A)**) and two-fold anisotropy in the crossed configuration (**Figure 1 (B)** and [16]) may also reflect possible breaking of the local site and/or time-reversal symmetry in AFM phase of $MnTe_2$.

Further investigation of the delicate mechanism of interplay between the magnetic spin and local atomic ordering in AFM phase of $MnTe_2$ is required.

## CONCLUSIONS.

In summary, we have investigated the influence of antiferromagnetic ordering on the lattice dynamics and low-energy magnetic excitations of pyrite-type $MnTe_2$ through comprehensive study of light scattering using Raman spectroscopy. Temperature-dependent measurements reveal pronounced deviations from conventional anharmonic behavior for four of the five Raman-active phonon modes below the Néel temperature. These deviations are strongly mode-dependent, including phonon hardening, softening, and nonmonotonic temperature evolution, whereas the fully symmetric $A_g$ mode follows the expected anharmonic behavior. The observed renormalization of phonon frequencies provides clear evidence of a strong interplay between lattice vibrations and magnetic ordering in $MnTe_2$. The experimental observations are qualitatively reproduced by a theoretical model incorporating spin-phonon coupling within a Green's-function formalism. Despite its simplified description of the magnetic structure, the model captures the main features of the phonon frequency renormalization associated with spin correlations. Additionally, magnetic-field-dependent Raman measurements reveal distinct responses of the two low-energy magnon modes, including Zeeman splitting of the higher-energy excitation and weak softening of the lower-energy mode. Moreover, we discuss breaking of the local site and/or time-reversal symmetry as a possible driving force enabling magnetoelastic interactions leading to the observed anomalous temperature-dependence of selected phonon modes as well as twofold anisotropy and peak shape asymmetry of the higher-energy magnon mode. These findings highlight the important role of spin-phonon interactions in determining the vibrational properties of $MnTe_2$ and provide a basis for further microscopic investigations of spin-lattice coupling in unconventional antiferromagnets.

## EXPERIMENTAL METHODS

### Sample description, synthesis and characterization.

Single crystalline samples of $MnTe_2$ were used for study. If otherwise was clearly stated, measurements were performed on the same sample piece.

Single crystals of $MnTe_2$ were grown using the chemical vapor transport (CVT) method. Stoichiometric amounts of high-purity manganese (Mn, 3N8, Koch-Light Laboratories) and tellurium (Te, 5N, Koch-Light Laboratories) were weighed and loaded into a quartz ampoule along with iodine (POCH, 99.8%) as the transport agent, in a concentration of 3 mg/cm³ of the ampoule volume. The ampoule was then evacuated and sealed under high vacuum.

The sealed ampoule was placed in a three-zone horizontal furnace to precisely control the temperature gradient and ensure a flat crystallization zone. The growth process was conducted with the source zone maintained at 600 °C and the crystallization zone at 540 °C. This temperature gradient was sustained for 14 days to facilitate the chemical vapor transport and crystal growth. After the growth period, the furnace was switched off and allowed to cool down to room temperature. The single crystals of $MnTe_2$ were verified by x-ray diffraction analysis showing pyrite type cubic crystal structure.

To reduce residual tellurium from the surface of the as-grown crystals, selected crystals were annealed under vacuum at 300 °C for 3 hours.

### Raman spectroscopy study

Raman scattering measurements were performed using three experimental setups Raman spectrometers: Raman spectrometers #1 [LabRAM Aramis (Horiba Jobin Yvon)] for initial characterization, spectrometer #2 [SpectraPro HRS 750 (Princeton Instruments)] for low temperature nonpolarized and polarized Raman measurements and #3 [SpectraPro HRS 500 (Princeton Instruments)] for measurements in high magnetic fields.

**Temperature-dependent measurements.**

Nonpolarized Raman scattering spectra were collected in backscattering geometry from the sample inside **the continuous-flow Cryovac Konti Micro cryostat in the temperature range from 5 to 300 K.** The sample was excited using a 632.8 nm of He-Ne laser (Thorlabs). The laser power was attenuated to ca. 500 µW. Incident laser beam was focused on the probed sample using 50x long working distance objective (Mitutoyo). Scattered light was collected by the same objective and transferred to Raman spectrometer #2 equipped with diffraction grating 1800 l/cm and **the liquid nitrogen-cooled** charge coupled device detector (CCD) [**PyLoN eXcelon**) (Princeton Instruments)] yielding spectral resolution ca. 0.5 $cm^{-1}$. Rayleigh scattering was attenuated by a set of diffracting filters (**Optigrate**). Stokes part of Raman spectra was measured in the range from ca. 10 to 200 $cm^{-1}$. Acquisition time was 100 s.

**Polarization-dependent measurements.**

Measurements were carried out using the same experimental setup as for temperature-dependent measurements**.** Raman spectra were measured in backscattering geometry with parallel ($\hat{e}_i \parallel \hat{e}_s$) and perpedincular ($\hat{e}_i \perp \hat{e}_s$) configurations, where $\hat{e}_i$ and $\hat{e}_s$ are polarization vectors of incident and scattered light, respectively. For the measurements the natural (100) face of the crystal was used.

The polarization of the detected Raman signal was controlled using a motorized half-wave plate placed in the detection path. The sample was excited using a 515-nm diode laser (Cobolt). The laser power was attenuated to ca. 500 µW. Stokes part of Raman spectra were measured in the range from 10 to 250 $cm^{-1}$. Acquisition time was 60 s.

**Excitation-dependent measurements.**

Nonpolarized Raman scattering spectra were collected from the same sample with crystal face (100) orthogonal with respect to the laser beam. Besides 632.8 nm excitation used for temperature-dependent measurements, the sample was excited by 515 nm laser line coupled with Raman spectrometers #2. Power of the probing lasers was kept well below 1 mW in order to avoid a noticeable buildup of the signal related to elementary Tellurium in the measured spectra.

**Magneto-optical measurements (Field-dependent Raman scattering)**

Nonpolarized Raman scattering spectra were collected in backscattering geometry from the sample inside the setup housing the superconducting Cryogenic magnet generating magnetic fields up to 16 T over a wide temperature range (6 - 300 K). Measurements were carried out in the Voigt configuration, where the external magnetic field was applied perpendicular to the direction of light propagation. The sample was placed in a specially designed probe with optical access, and a set of Attocube piezo elements enables spatial mapping. The focusing of the excitation light as well as the collection of the signal was ensured by the Attocube microscope objective, which could operate at low temperatures and in magnetic fields, providing an excitation spot of about 1 $\mu m^2$. Scattered light was collected by the same objective and transferred to Raman spectrometer #3 equipped with diffraction grating 1800 l/cm and the electrically-cooled CCD [PIXIS eXcelon (Princeton Instruments)] yielding spectral resolution ca. 0.5 $cm^{-1}$. Rayleigh scattering was attenuated by a set of diffracting filters (Optigrate). Stokes part of Raman spectra was measured in the range from ca. 10 to 200 $cm^{-1}$. Acquisition time was 100 s.

**ACKNOWLEDGEMENTS**

A.S.W., A.W. and D. J. acknowledge the support of the “MagTop” project (FENG.02.01-IP.05-0028/23) carried out within the “International Research Agendas” programme of the Foundation for Polish Science, co-financed by the European Union under the European Funds for Smart Economy 2021-2027 (FENG). A.S.W. acknowledges the support of the National Science Centre, Poland (NCN), through the MINIATURA 9 with project No. 2025/09/X/ST3/00809. E. Z. gratefully acknowledges the support of the IEEE Magnetics Society (Magnetism for Ukraine Initiative program) via the STCU Project No. 9918. M.R.M. acknowledges the support of the National Science Centre, Poland (grant no. 2023/50/O/ST3/00310).

**AUTHOR CONTRIBUTIONS:**
**Taras Palasyuk:** conceptualization, methodology, investigation, formal analysis, data curation, project administration, resources, supervision, validation, visualization, writing – original draft, writing – review & editing
**Ashutosh S. Wadge:** conceptualization, single crystals of $MnTe_2$, writing – review & editing
**Eduard Zubov:** investigation, formal analysis, data curation, writing – original draft, writing – review & editing
**Igor Antoniazzi:** investigation, formal analysis, resources, writing – review & editing
**Gayatri:** investigation, writing – review & editing
**Andrzej Wiśniewski:** conceptualization, resources, writing – review & editing
**Wojciech Szuszkiewicz:** writing – review & editing
**Daniel Jastrzebski:** resources, writing – review & editing
**Ryszard Diduszko:** resources, writing – review & editing
**Adam Babiński:** resources, writing – review & editing
**Maciej R. Molas:** resources, writing – review & editing
**Michał Wierzbicki:** formal analysis
**Cezariusz Jastrzebski:** conceptualization, methodology, resources, investigation, formal analysis, data curation, writing – review & editing

**REFERENCES.**